\documentclass{article}

\usepackage{spconf}
\usepackage{amsmath}
\usepackage{eso-pic,rotating,graphicx}
\usepackage{pifont}
\usepackage{xcolor}
\usepackage{caption}
\usepackage{subcaption}
\usepackage{algorithm}
\usepackage{algpseudocode}
\usepackage[algo2e, ruled,vlined]{algorithm2e}
\usepackage{hyperref}
\usepackage{cite}
\usepackage{adjustbox}
\usepackage{multirow}
\usepackage{color, colortbl}
\definecolor{Gray}{gray}{0.9}

\newcommand{\cmark}{\ding{51}}%
\newcommand{\xmark}{\ding{55}}%

\AddToShipoutPicture{\put(590,750){\rotatebox{270}{\scalebox{0.6}{Copyright 2023 IEEE. Published in the 2022 IEEE Spoken Language Technology Workshop (SLT) (SLT 2022), scheduled for 19-22 January 2023 in Doha, Qatar. Personal use of this material is permitted. However, permission for advertising or promotional purposes or  for creating new collective works  for resale or}}}}
\AddToShipoutPicture{\put(585,750){\rotatebox{270}{\scalebox{0.6}{redistribution to servers or lists, or to reuse any copyrighted component of this work  in other works, must be obtained from the IEEE. Contact: Manager, Copyrights and Permissions / IEEE Service Center / 445 Hoes Lane / P.O. Box 1331 / Piscataway, NJ 08855-1331, USA. Telephone: + Intl. 908-562-3966.}}}}

\title{Low-Latency Speech Separation Guided Diarization\\
for Telephone Conversations}

\name{%
\begin{tabular}{@{}c@{}}
Giovanni Morrone$^{*1}$, Samuele Cornell$^{*1}$, Desh Raj$^2$, Luca Serafini$^1$ \\ 
Enrico Zovato$^3$, Alessio Brutti$^4$, Stefano Squartini$^1$ \thanks{$^*$ denotes equal contribution.}
\end{tabular}}
 \address{
   $^1$Università Politecnica delle Marche, Ancona, Italy
   $^2$Johns Hopkins University, Baltimore, USA \\
   $^3$PerVoice S.p.A., Trento, Italy
   $^4$Fondazione Bruno Kessler, Trento, Italy}

\copyrightnotice{978-1-6654-7189-3/22/\$31.00~\copyright2023 IEEE}

\begin{document}
\ninept

\maketitle

\begin{abstract}
% Length: from 100 to 150 words
% Max 1000 characters
%Speech separation and speaker diarization have strong similarities especially considering end-to-end neural diarization (EEND) methods.
In this paper, we carry out an analysis on the use of speech separation guided diarization (SSGD) in telephone conversations. SSGD performs diarization by separating the speakers signals and then applying voice activity detection on each estimated speaker signal. In particular, we compare two low-latency speech separation models. 
Moreover, we show a post-processing algorithm that significantly reduces the false alarm errors of a SSGD pipeline. 
We perform our experiments on two datasets: Fisher Corpus Part 1 and CALLHOME, evaluating both separation and diarization metrics.
Notably, our SSGD DPRNN-based online model achieves $11.1$\% DER on CALLHOME, comparable with most state-of-the-art end-to-end neural diarization models despite being trained on an order of magnitude less data and having considerably lower latency, i.e., $0.1$ vs. $10$ seconds.
We also show that the separated signals can be readily fed to a speech recognition back-end with performance close to the oracle source signals. 

%Separation aims at extracting each speaker from overlapped speech, while diarization identifies time boundaries of speech segments produced by the same speaker.
%In particular with respect to end-to-end neural diarization (EEND) methods.
%, both in offline and online settings. 
%In the online setting we consider both the use of continuous source separation (CSS) and causal SSep models architectures. 
%We focus on two-speaker real-world telephone conversations scenarios and perform our experiments on Fisher Corpus Part 1 and CALLHOME datasets evaluating both separation and diarization metrics.

\end{abstract}

\section{Introduction}
\label{sec:intro}

Speaker diarization (or ``who spoke when'') is the task of segmenting a recording into homogeneous speaker-specific regions~\cite{anguera2012speaker,park2022review}. It constitutes an important preprocessing step for many applications, such as meeting summary, live captioning, speaker-based indexing, and telephone conversation analysis.

Diarization methods can be broadly divided into two categories: clustering-based and end-to-end supervised systems.
The former typically employs a pipeline comprised of voice activity detection (VAD), speaker embedding extraction and clustering \cite{garcia2017speaker, wang2018speaker, park2019auto, landini2022bayesian}. End-to-end neural diarization (EEND) reformulates the task as a multi-label classification. The majority of these systems \cite{fujita2019end, fujita2020end, horiguchi2020end} are trained directly to perform diarization using permutation invariant training (PIT) \cite{kolbaek2017multitalker}. There also exist methods such as target-speaker VAD~\cite{Medennikov2020TargetSpeakerVA} and region proposal networks~\cite{huang2020speaker} which lie at the intersection of these two categories.

With enough training data, end-to-end approaches have been shown to outperform state-of-the-art clustering-based systems~\cite{Horiguchi2021TheHD}, but at the cost of requiring significant memory for long recordings (e.g., longer than $10$ minutes). Chunk-wise processing can help reduce the memory footprint but it leads to inter-window speaker label permutation problem due to the PIT training objective \cite{xue2021online, han2021bw, xue2021edaonline, kinoshita2021integrating}.
Several recent approaches have been proposed to address this problem, such as employing a speaker tracing buffer \cite{xue2021online, xue2021edaonline}, a hybrid end-to-end/clustering framework \cite{kinoshita2021integrating} or chunk-level recurrence in the chunk hidden states \cite{han2021bw}. %In \cite{han2021bw} a chunk-level recurrence has exploited to process the chunk hidden states making the model complexity linear in time.
Another system \cite{zeghidour2021dive} iteratively builds embeddings for each speaker which are exploited to condition the following VAD module. 
However, most of EEND methods, with the exception of \cite{xue2021online, han2021bw, xue2021edaonline}, work offline and are thus not suitable for streaming applications such as live captioning. 

A key reason for the adoption of end-to-end diarization is its advantage over clustering-based methods in handling overlapped speech. Conventional clustering algorithms inherently assume single-speaker segments, and are thus prone to missing out on overlapping speakers (which may constitute as high as 20\% of the speech in real conversations \cite{watanabe2020chime}). Although researchers have proposed techniques for overlap assignment in VBx~\cite{bullock2020overlap} and spectral clustering~\cite{raj2021multi}, these methods depend heavily on an accurate overlap detection, which is often challenging to train. Furthermore, embedding extractors trained on single-speaker utterances may not produce reliable representations for overlapping segments, resulting in speaker confusion errors in these regions~\cite{raj2021multi}.

An alternative framework to deal with overlapped speech is continuous speech separation (CSS)~\cite{chen2020continuous, morrone2022conversational}.
CSS extends PIT-based speech separation (SSep) to long recording scenarios, by applying separation in a chunk-wise manner, where each chunk is assumed to contain a fixed number of speakers (usually 2-3). Since the underlying separator is trained via a PIT objective, output permutation consistency between chunks is not guaranteed. CSS solves this problem by performing overlapping inference (i.e., using strides shorter than chunk sizes) and reordering adjacent chunks based on a similarity measure over the portion in which they overlap.  
Several recent works have proposed diarization systems based on CSS by applying clustering techniques across the separated audio streams~\cite{raj2021integration, Xiao2021MicrosoftSD}. Fang et al.~\cite{fang2021deep} have proposed speech separation guided diarization (SSGD), where diarization is performed by first separating the input mixture and applying a conventional VAD to detect speech segments in each channel. SSGD is a particularly appealing approach as separated sources could be readily used for downstream tasks such as automatic speech recognition (ASR), with diarization coming almost ``for free''. We show this in Section \ref{subsec:res/wer_eval}. 

In this work, we build upon the SSGD framework and attempt to deal with its limitations. We extend SSGD to online processing by considering the use of causal SSep models and CSS. Both these techniques allow the processing of arbitrarily long recordings that could not fit in memory making SSGD viable in practical applications. Additionally, we introduce an effective, causal leakage removal post-processing algorithm that reduces the SSGD VAD false alarm errors generated by imperfect separation.
This algorithm has negligible computational overhead. 
We carry out an extensive experimental analysis focusing on conversational telephone speech (CTS). Although the maximum number of speakers is limited to $2$, this scenario is very common in many commercial applications that deal with CTS processing, and, in general, with analysis of conversations between two speakers (e.g., doctor-patient recordings).
The CTS scenario also allows to compare with most previous works \cite{fujita2019end, fujita2020neural, fujita2020end, horiguchi2020end, kinoshita2021integrating, zeghidour2021dive, xue2021online, han2021bw, xue2021edaonline}, on EEND diarization.
We experiment with real-world CTS datasets such as Fisher \cite{cieri2004fisher} and CALLHOME \cite{callhome}, comparing several online approaches and the effect of the CSS window size on the diarization accuracy.
In this preliminary work, we do not consider other datasets (e.g., CHiME \cite{watanabe2020chime}, AMI \cite{carletta2005ami} and DIHARD \cite{Ryant2019}) used in previous work as they do not belong to the CTS domain. %We will consider them in future work.
% comparing offline and online approaches and studying the effect of the CSS window size.
Results show that by using just separation and a simple VAD, it is possible to obtain competitive diarization results on CALLHOME with extremely low-latency (i.e., $0.1$ vs $10$ s) and using much less training data (i.e., ${\sim} 900$ vs ${\sim} 10000$ hours) compared to state-of-the-art EEND approaches.
Our code is made available at \url{https://github.com/dr-pato/SSGD}. %\footnote{The code will be made available upon acceptance.}.

\section{System Description}
\label{sec:sys}

\begin{figure}[t]
\centering
\includegraphics[width=0.45\textwidth]{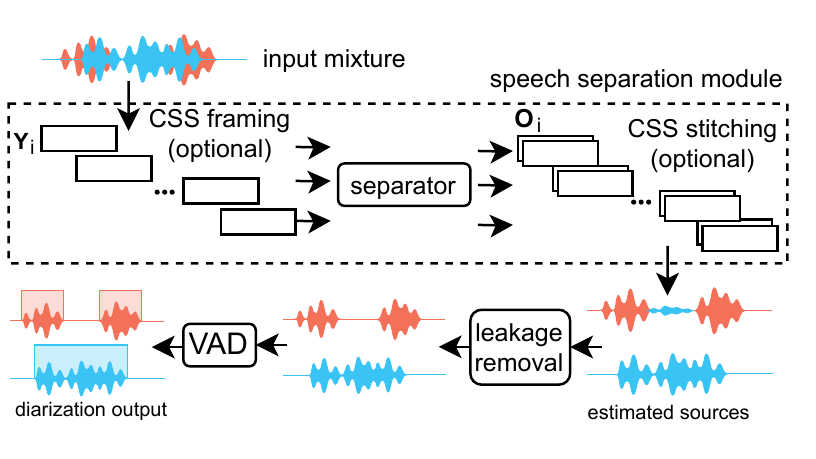}
\vspace{-0.6cm}
\caption{General diagram for the SSGD method.}
\label{fig:cssgd}
\end{figure}

\ClearShipoutPicture

The adopted SSGD pipeline is composed of three modules: speech separation, leakage removal post-processing and VAD, as shown in Fig. \ref{fig:cssgd}. The input of the system is a single-channel mixed audio stream, denoted as $\mathbf{Y} \in \mathbb{R}^{1 \times T}$, where $T$ is the number of audio samples.

\subsection{Speech Separation Module}
\label{subsec:sys/css}
%As said, we consider in our experiments SSGD based on non-causal separation models (as used in \cite{fang2021deep}), causal separation models (such as causal Conv-TasNet \cite{luo2019conv}) and CSS. 
We consider in our experiments SSGD based on causal separation models (i.e., Conv-TasNet \cite{luo2019conv} and DPRNN \cite{luo2020dual}).
Since the majority of diarization approaches only work offline, we also experiment with non-causal separation models (as used in \cite{fang2021deep}) to carry out a more comprehensive comparison with clustering-based and EEND state-of-the-art systems.
Additionally, we analyze the application of CSS with non-causal SSep models. In such configuration, the latency of these models is tied to the CSS window size and thus can be used online. CSS is not applied to causal SSep models since they are already capable to process the input in a streaming fashion with low latency. 

Briefly, CSS consists of three stages as shown in Fig.~\ref{fig:cssgd}: framing, separation and stitching. In the framing stage, a windowing operation splits $\mathbf{Y}$ into $I$ overlapped frames $\mathbf{Y}_i \in \mathbb{R}^{1 \times W}, i = 1, \dots, I$, with $I = \lceil \frac{T}{H} \rceil$, where $W$ and $H$ are the window and hop sizes, respectively. %Window size defines the memory footprint and the latency of the system. 
Then, the separation is performed independently on each frame $\mathbf{Y}_i$, generating separated output frames $\mathbf{O}_i \in \mathbb{R}^{C \times W}$, where $C$ is the number of output channels. In this work, $C$ is fixed to $2$, meaning that we assume that the maximum number of speakers in any frame is $2$. This is a common assumption made for CSS systems, and is also valid in general for telephone conversations (which is the focus of this work). To solve the permutation ambiguity between consecutive frame outputs, the stitching module aligns channels of two separation outputs $\mathbf{O}_i$ and $\mathbf{O}_{i+1}$ according to the cross-correlation computed on the overlapped part of consecutive frames. The final output stream $\mathbf{X} \in \mathbb{R}^{C \times T}$ is generated by an overlap-add operation with an Hanning window.

% In this preliminary work, we consider the application of CSS only to non-causal SSep models, which are generally more performant than their causal counterpart \cite{luo2019conv}. In such configuration, the latency of these models will be tied to the CSS stride and thus can be used online.
% On the other hand, causal separation models are already capable to process the input in a streaming fashion with small latency. 

\vspace{-0.2cm}

\subsection{Leakage Removal Post-Processing}
\label{subsec:sys/post_proc}

In the presence of long input recordings, even state-of-the-art separation models are prone to channel leakage when only one speaker is active (e.g., see \textit{estimated sources} in Fig. \ref{fig:cssgd}). As a result, the ``leaked'' segments are detected as speech by the following VAD module, leading to a higher false alarm error in the final diarization output. To alleviate this problem, we propose a post-processing algorithm to reduce false alarms without significantly affecting missed speech, speaker confusion errors, and separation quality. It does not introduce additional latency and its computational overhead is negligible.

Given an input mixture $\mathbf{Y}$ and two estimated sources $\mathbf{X}^1$ and $\mathbf{X}^2$, we split each signal into disjoint segments $\mathbf{Y}_{\ell}$, $\mathbf{X}_{\ell}^1$, $\mathbf{X}_{\ell}^2$ of length $L$. For each segment, we compute the Scale-Invariant Signal-to-Distortion Ratios (SI-SDR) \cite{leroux2019sdr} $s_{\ell}^1$, $s_{\ell}^2$ between segments of every source $\mathbf{X}_{\ell}^1$, $\mathbf{X}_{\ell}^2$ with the associated segment $\mathbf{Y}_{\ell}$ of input mixture. If both $s_{\ell}^1$, $s_{\ell}^2$ are above a threshold $t_{\ell r}$, a segment with leakage is detected. Leakage is removed by filling with zeros the segment with lower SI-SDR. This process results in new estimated sources $\mathbf{\tilde{X}_{\ell}}$, which are passed as input to the VAD module. The leakage removal algorithm is summarized in the pseudocode below.

\vspace{-3pt}

\SetKwInput{KwInput}{Input}                
\SetKwInput{KwOutput}{Output}
\SetKw{KwBy}{by}
\SetKw{KwAnd}{and}
\begin{algorithm}[H]
\DontPrintSemicolon
  \KwInput{$\mathbf{Y}$, $\mathbf{X}^1$, $\mathbf{X}^2$, $T$, $L$, $t_{\ell r}$}
  \KwOutput{$\mathbf{\tilde{X}_{\ell}}^1$, $\mathbf{\tilde{X}_{\ell}}^2$}
  
  $\mathbf{\tilde{X}_{\ell}}^1 \gets \mathbf{X}^1$; $\mathbf{\tilde{X}_{\ell}}^2 \gets \mathbf{X}^2$ 
  
  %$N = \lfloor \frac{T}{L} \rfloor$
  
  \For{$i \gets 0$ \KwTo $T$ \KwBy $L$}{
    $s_{\ell}^1 \gets $ SI-SDR$(\mathbf{Y}$[$i$:$i$+$L$], $\mathbf{X}^1$[$i$:$i$+$L$])
    
    $s_{\ell}^2 \gets $ SI-SDR$(\mathbf{Y}$[$i$:$i$+$L$], $\mathbf{X}^2$[$i$:$i$+$L$])
    
    \If{ $s_{\ell}^1 > t_{\ell r}$ \KwAnd $s_{\ell}^2 > t_{\ell r}$}{
      \If{$s_{\ell}^1 > s_{\ell}^2$}{
        $\mathbf{\tilde{X}_{\ell}}^2$[$i$:$i$+$L$] $ \gets 0$
      }
      \Else{
       $\mathbf{\tilde{X}_{\ell}}^1$[$i$:$i$+$L$] $ \gets 0$
      }
    }
    }
\caption{Leakage Removal}
\end{algorithm}

\vspace{-0.65cm}

\subsection{Voice Activity Detection (VAD)}
\label{subsec:sys/vad}

The VAD module is used to extract active speech segments from the post-processed estimated sources and generate the diarization output. It is applied on each estimated source $\mathbf{\tilde{X}_{\ell}}$ independently but future work could also consider a multi-source VAD.
We experiment with two different VAD models: an energy-based VAD ~\cite{landini2021analysis}, and a neural model which employs a temporal convolutional network (TCN), as proposed in \cite{cornell2022overlapped}.

\begin{table*}[t]
\centering
\caption{Speech separation and diarization results on the Fisher and CALLHOME test sets in the \textbf{online} scenario. Separation is assessed using the SI-SDR (dB) improvements over the input mixtures. Diarization is assessed using diarization error rate (DER), missed speech (MS), false alarm (FA) and speaker confusion errors (SC). Latency of the system is reported in seconds.
%In addition to SI-SDRi and DER metrics, we also show latency of the systems (in seconds).
The best results among proposed techniques are shown in \textbf{bold}, and among EEND methods are \underline{underlined}.}
\vspace{3pt}

\label{tab:res/diar_real_online}
\adjustbox{max width=\textwidth}{%
\centering
\begin{tabular}{@{}llcccccccccc@{}}
\toprule
\multirow{2}{*}{\textbf{Method}} & \multirow{2}{*}{\textbf{VAD}} & \multicolumn{1}{l}{\multirow{2}{*}{\textbf{Latency (s)}}} & \multicolumn{5}{c}{\textbf{Fisher}} & \multicolumn{4}{c}{\textbf{CALLHOME}} \\
\cmidrule(r{4pt}){4-8} \cmidrule{9-12}
 &  & \multicolumn{1}{l}{} & \multicolumn{1}{l}{\textbf{SI-SDRi}} & \multicolumn{1}{l}{\textbf{MS}} & \multicolumn{1}{l}{\textbf{FA}} & \multicolumn{1}{l}{\textbf{SC}} & \multicolumn{1}{l}{\textbf{DER}} & \multicolumn{1}{l}{\textbf{MS}} & \multicolumn{1}{l}{\textbf{FA}} & \multicolumn{1}{l}{\textbf{SC}} & \multicolumn{1}{l}{\textbf{DER}} \\
\midrule
SA-EEND w/STB~\cite{xue2021online} & & 1 &  &  &  &  &  &  &  &  & 12.5 \\
BW-EDA-EEND~\cite{han2021bw} & & 10 &  &  &  &  &  &  &  &  & 11.8 \\
SA-EEND-EDA w/STB~\cite{xue2021edaonline} & & 10 &  &  &  &  &  &  &  &  & \underline{10.0} \\
\midrule
%(Offline) DPRNN + Leakage removal & TCN & & 21.10 & 4.2 & 1.4 & 1.4 & 7.0 & 6.0 & 2.8 & 0.9 & 9.7 \\
%\midrule
\textit{Oracle sources} & Energy & & $\infty$ & 7.4 & 1.4 & 0.1 & 8.9 &  &  &  &  \\
\textit{Oracle sources} & TCN & & $\infty$ & 3.2 & 1.7 & 0.1 & 5.0 &  &  &  &  \\
Conv-TasNet & Energy & 0.01 & -0.9 & 11.5 & 39.1 & 9.5 & 60.1 & 7.3 & 55.8 & 5.6 & 68.7 \\
Conv-TasNet & TCN & 0.01 & -0.9 & 1.7 & 70.3 & 2.2 & 74.1 & \textbf{3.4} & 82.3 & 0.6 & 86.4 \\
~~ + Leakage removal & TCN & 0.01 & -3.1 & 5.2 & 5.6 & 25.9 & 36.8 & 6.2 & 21.9 & 15.5 & 42.6 \\
%DPRNN (v1) & Energy & 0.1 & 19.8 & 7.9 & 4.7 & 0.9 & 13.4 & \textbf{5.7} & 12.8 & 1.3 & 19.8 \\
%DPRNN (v1) & TCN & 0.1 & 19.8 & 5.7 & 6.2 & \textbf{0.2} & 12.1 & 5.9 & 13.4 & 1.0 & 20.2 \\
%~~ + Leakage removal & TCN & 0.1 & 19.2 & 4.9 & \textbf{1.2} & 1.7 & 7.9 & 6.5 & 3.7 & 2.5 & 12.7 \\
DPRNN & Energy & 0.1 & \textbf{22.6} & 7.6 & \textbf{1.4} & \textbf{0.8} & 9.7 & 5.5 & 6.9 & 1.9 & 14.3 \\
DPRNN & TCN & 0.1 & \textbf{22.6} & \textbf{3.8} & 2.6 & \textbf{0.8} & 7.1 & 5.9 & 4.5 & \textbf{1.6} & 12.0 \\
~~ + Leakage removal & TCN & 0.1 & 22.2 & 4.3 & 1.8 & \textbf{0.8} & \textbf{6.8} & 6.9 & \textbf{2.3} & 1.9 & \textbf{11.1} \\
\bottomrule
\end{tabular}
}
\vspace{-0.5cm}
\end{table*}

\section{Experimental Setup}
\label{sec:exp_setup}

\subsection{Datasets}
\label{ssec:exp_setup/dataset}

Since the focus of our work is on the CTS scenario, we use the \emph{Fisher Corpus Part 1} \cite{cieri2004fisher} for both training and test purposes.
Fisher consists of 5850 telephone conversations in English, sampled at 8 kHz, between two participants. It provides a separated signal for each of the two speakers. This allows training a separation model directly on this dataset and computing source separation metrics such as the SI-SDR improvement (SI-SDRi) \cite{leroux2019sdr}. 
Training, validation and test sets are created by drawing $5728$, $61$, and $61$ conversations, respectively, with no overlap between speakers identities. The amount of overlapped speech is around 14\% of the total speech duration. 
In addition, we generate a simulated fully-overlapped version of Fisher for the purpose of training the SSep models. This portion is derived from the training set and amounts to 30000 mixtures for a total of 44 hours.  

We also test the proposed methods on the portion of the 2000 NIST SRE ~\cite{callhome, slr10} denoted as \emph{CALLHOME}, consisting of real-world multilingual telephone conversations. Following the recipe in \cite{fujita2019end}, we use the 2-speaker subset of CALLHOME and the adaptation/test split that allows to compare with most end-to-end diarization methods mentioned previously. 
The amount of overlapped speech is around 13\% of total speech duration.

\begin{table*}[t]
\centering
\caption{Speech separation and diarization results on the Fisher and CALLHOME test sets in the \textbf{offline} scenario. The best results among proposed techniques are shown in \textbf{bold}, and those among baselines are \underline{underlined}. \textit{Oracle sources} evaluation is the same of Table \ref{tab:res/diar_real_online}, as the VADs works online in both online and offline scenarios.}
\vspace{3pt}

\label{tab:res/diar_real}
\adjustbox{max width=\textwidth}{%
\centering
\begin{tabular}{@{}llccccccccc@{}}
\toprule
\multirow{2}{*}{\textbf{Method}} & \multirow{2}{*}{\textbf{VAD}} &  \multicolumn{5}{c}{\textbf{Fisher}} & \multicolumn{4}{c}{\textbf{CALLHOME}} \\
\cmidrule(r{4pt}){3-7} \cmidrule{8-11}
 &  & \multicolumn{1}{l}{\textbf{SI-SDRi}} & \multicolumn{1}{l}{\textbf{MS}} & \multicolumn{1}{l}{\textbf{FA}} & \multicolumn{1}{l}{\textbf{SC}} & \multicolumn{1}{l}{\textbf{DER}} & \multicolumn{1}{l}{\textbf{MS}} & \multicolumn{1}{l}{\textbf{FA}} & \multicolumn{1}{l}{\textbf{SC}} & \multicolumn{1}{l}{\textbf{DER}} \\
\midrule

VBx~\cite{landini2022bayesian} & TCN &  & 10.0 & \underline{0.3} & 0.5 & 10.8 & 7.3 & 1.9 & 3.1 & 12.3 \\
VBx~\cite{landini2022bayesian} & Kaldi &   & 8.9 & 0.4 & 0.9 & 10.2 & 8.3 & \underline{0.9} & 2.6 & 11.7 \\
~~ + Overlap assignment~\cite{bullock2020overlap} & Kaldi &   & \underline{4.4} & 2.1 & 0.9 & \underline{7.4} & 5.3 & 2.5 & 2.4 & 10.3 \\
Spectral clustering~\cite{park2019auto} & Kaldi &    & 8.9 & 0.4 & \underline{0.2} & 9.5 & 8.3 & \underline{0.9} & 5.3 & 14.5 \\
~~ + Overlap assignment~\cite{raj2021multi} & Kaldi &   & 5.2 & 2.0 & \underline{0.2} & \underline{7.4} & 5.7 & 2.7 & 5.8 & 14.1 \\
SA-EEND~\cite{fujita2019end} &  &  &    &  &  &  &  &  &  & 9.5 \\
SA-EEND-EDA~\cite{horiguchi2020end} &    &  &  &  &  &  &  &  &  & 8.1 \\
EEND + VC~\cite{kinoshita2021integrating} &  &    &  &  &  &  & \underline{4.0} & 2.4 & \underline{0.5} & 7.0 \\
DIVE~\cite{zeghidour2021dive} &  &  &  &   &  &  &  &  &  & \underline{6.7} \\
\midrule
% \textit{Oracle sources} & Energy   & $\infty$ & 7.4 & 1.4 & 1.0 & 8.9 &  &  &  &  \\
% \textit{Oracle sources} & TCN  & $\infty$ & 3.2 & 1.7 & 0.1 & 5.0 &  &  &  &  \\
Conv-TasNet & Energy  & 17.5 & 8.0 & 4.5 & 1.6 & 14.1 & 6.0 & 12.0 & 2.8 & 20.6 \\
Conv-TasNet & TCN  & 17.5 & 6.2 & 5.0 & 1.1 & 12.4 & 6.1 & 13.6 & 1.8 & 21.6 \\
~~ + Leakage removal & TCN  & 17.1 & 5.5 & 2.5 & 2.0 & 10.1 & 6.0 & 10.1 & 2.8 & 18.9 \\
%DPRNN (v1) & Energy & \textbf{21.3} & 7.8 & 2.3 & 1.0 & 11.1 & 5.4 & 5.9 & 0.6 & 12.0 \\
%DPRNN (v1) & TCN  & 21.3 & 4.1 & 3.1 & 0.9 & 8.1 & 4.8 & 9.0 & 0.2 & 14.1 \\
%~~ + Leakage removal & TCN  & 21.1 & 4.2 & 1.4 & 1.4 & 7.0 & 6.0 & 2.8 & 0.9 & 9.7 \\
DPRNN & Energy  & \textbf{22.6} & 7.6 & \textbf{1.2} & \textbf{0.7} & 9.5 & 5.5 & 4.4 & 0.5 & 10.4 \\
DPRNN & TCN  & \textbf{22.6} & \textbf{3.4} & 2.2 & \textbf{0.7} & 6.3 & \textbf{5.0} & 5.4 & \textbf{0.4} & 10.8 \\
~~ + Leakage removal & TCN  & 22.2 & 3.9 & 1.6 & \textbf{0.7} & \textbf{6.1} & 6.6 & \textbf{1.9} & 0.7 & \textbf{9.3} \\
\bottomrule
\end{tabular}
}
\vspace{-0.5cm}
\end{table*}

\vspace{-0.2cm}

\subsection{Architecture, Training and Inference Details}

We employ the Asteroid toolkit \cite{pariente20} to experiment with $2$ SSep architectures: Conv-TasNet and DPRNN, both in online (causal) and offline (non-causal) configurations (for a total of $4$).
For both, we use the best hyperparameter configuration as found in \cite{luo2019conv, luo2020dual} with these exceptions: to reduce memory footprint we employ a $16$ analysis/synthesis kernel size for encoder/decoder also for DPRNN and, regarding causal models, we use standard layer normalization versus the non-causal global layer normalization employed in non-causal models. Additionally, we set the DPRNN chunk and hop sizes to $100$ and $50$, respectively.
These models are trained on the simulated fully overlapped Fisher dataset %described in Section \ref{sec:fisher} 
using the SI-SDR objective to separate two speakers.
We use Adam optimizer \cite{kingma2015adam}, batch size $4$ and learning rate $0.001$. We clip gradients with $l_2$ norm greater than 5. Learning rate is halved if SI-SDR does not improve on validation for $10$ epochs. If no improvement is observed for $20$ epochs, training is stopped.
Each SSep model is then fine-tuned using a learning rate of $0.0001$ and batch size $1$ on the real Fisher data, by taking $60$ s long random segments from each recording.

We adopt the TCN VAD from \cite{cornell2022overlapped}, which is causal and for which the latency amounts to 10\,ms. This model is trained on the original Fisher data, using each speaker source separately, as the VAD is then applied to estimated sources. 
We train on random $2$ s long segments with a batch size of 256. The rest of training hyperparameters are the same as those used for SSep models. 
In inference we employ a median filter to smooth the VAD predictions. In addition, we remove segments shorter than a threshold $t_s$ to further reduce false alarm errors.
For each SSGD model, we tune the median filter, leakage removal threshold and $t_s$ parameters on the Fisher validation set (CALLHOME adaptation set for CALLHOME models).
The segment length $L$ of the leakage removal algorithm is set to 10\,ms, which results in the same latency as the TCN VAD.

% \subsection{Baseline Offline Methods}
% \label{subsec:res/baselines}

\section{Results}
\label{sec:res}

We evaluate the performance on Fisher and CALLHOME test sets in terms of diarization error rate (DER) including overlapped speech and using a collar tolerance of $0.25$ s, as in \cite{fujita2019end}. The evaluation is carried out using the standard NIST \textit{md-eval} scoring tool \cite{mdeval}.
For the Fisher test set we also report the SI-SDRi \cite{leroux2019sdr} source separation metric since oracle sources are available.

\vspace{-0.2cm}

\subsection{Online Separation/Diarization}
\label{subsec:res/online_eval}
The results for online SSGD diarization models are reported in Table~\ref{tab:res/diar_real_online}. Oracle sources refers to SSGD with oracle SSep, thus with error coming only from the VAD module. We carry out the oracle evaluation only for Fisher, as for CALLHOME separated sources are not provided. For the CALLHOME evaluation, we also show DERs obtained by EEND, as reported in the original papers.

We observed that the Conv-TasNet model failed to deal with long recordings, generating large false alarm errors. %We found that the model generated large false alarms that could be only partially mitigated by the leakage removal algorithm.
This is due to the fact that, being fully convolutional, it has a limited ${\sim} 1.5$ s receptive field. On the other hand, the DPRNN, being based on recurrent neural networks, has no such limitations and was effectively able to track the speakers for much longer and generate better diarization results.
The proposed leakage removal algorithm was highly effective for both architectures. This was especially true in the case of TCN-based VAD since it was more prone to false alarms caused by leaked speech due to being trained on real Fisher data and not on the output of the separators. Although the algorithm was only partially able to mitigate the low separation capability of the Conv-TasNet, it improved the DER by 50.3\% and 50.7\% on Fisher and CALLHOME, respectively. For DPRNN, the improvement was lower as the system without leakage removal was already able to obtain good diarization performance. However, the proposed post-processing almost halved the false alarm error rates and improved the DER by 4.2\% and 7.5\% on Fisher and CALLHOME, respectively.

As a comparison, the current best performing online system on the CALLHOME dataset (i.e., SA-EEND-EDA with speaker tracing buffer \cite{xue2021edaonline}), obtains 10.0\% DER, which is slightly better than ours but is obtained with significantly higher latency of $10$ s. Our approach works with a latency of $0.1$ s, making it appealing for applications where real-time requirements are very important (e.g., real-time captioning).
Last but not least, the SSGD is trained using a dataset of ${\sim} 900$ hours of speech, which is considerably smaller than the ones used to train the state-of-the-art EEND models (i.e., ${\sim} 10000$ hours) and results in shorter training times and less burden regarding additional costs for the generation of simulated mixtures. %Note that EEND systems are usually trained with simulated mixtures, then they could use an arbitrary amount of training data. However, the generation of simulated mixtures does not come for free as it introduces additional costs (e.g. computational time and disk space) compared to using original real recordings directly.

% As a result, DPRNN continued to be effective in the online scenario. Similar to the offline case, leakage removal added significant gains, improving the DER by 34.7\% and 37.1\% respectively on Fisher and CALLHOME, compared to the system without leakage removal. As a comparison, online variations of SA-EEND and SA-EEND-EDA using speaker tracing buffer~\cite{xue2021online,xue2021edaonline} obtain DERs 12.5\% and 10.0\% respectively, but with significantly higher latency of $1$ and $10$ s.

\vspace{-0.2cm}

\subsection{Offline Separation/Diarization}
\label{subsec:res/offline_eval}
For the offline scenario, we compare our approach with clustering-based and EEND methods. For the former, we use VBx~\cite{landini2022bayesian} and spectral clustering~\cite{park2019auto}, along with their overlap-aware counterparts \cite{bullock2020overlap, raj2021multi}. For VAD in these systems, we use the publicly available Kaldi ASpIRE VAD model~\cite{peddinti2015jhu}\footnote{\url{https://kaldi-asr.org/models/m4}}. For overlap detection, we fine-tune the Pyannote~\cite{bredin2020pyannote} segmentation model\footnote{\url{https://huggingface.co/pyannote/segmentation}} on the full CALLHOME adaptation set. The hyperparameters for each task are tuned on the corresponding validation set. The scripts for reproducing the baseline results are publicly available \footnote{\url{https://github.com/desh2608/diarizer}}. For fair comparison, we also report the performance of VBx with the TCN VAD, which however leads to degraded performance for this system.

The results for baselines and the offline SSGD diarization models are reported in Table~\ref{tab:res/diar_real}. 
As in Table \ref{tab:res/diar_real_online}, we show DERs of the EEND methods for the CALLHOME test set.

In contrast to the online scenario, Conv-TasNet obtained good separation capability. However, DPRNN-based SSGD strongly outperformed the Conv-TasNet version on all metrics on the Fisher dataset, and even surpassed the overlap-aware VBx which scored best among all clustering baselines.
Regarding separation performance (SI-SDRi), we can see that the offline DPRNN did not improve over the online one.
% On the Fisher dataset, DPRNN-based SSGD strongly outperformed the Conv-TasNet version on all metrics, and even surpassed the overlap-aware VBx which fares best among all clustering baselines.
In general, the TCN VAD outperformed the energy-based one, especially when the former was used jointly with the proposed leakage removal, which continued to be effective in the offline configuration.

For the CALLHOME data, the best performing SSGD model is comparable with SA-EEND \cite{fujita2019end}. Although the diarization capability is good, it is not competitive with the current best performing approaches \cite{kinoshita2021integrating, zeghidour2021dive}, making it less attractive for offline applications. 
However, as we show in Section \ref{subsec:res/wer_eval}, it can be a more cost effective solution as the separated signals can be readily used in downstream applications such as ASR.

In future work we will consider several strategies to reduce this gap such as training with more data, comparable to the amount used in EEND models, and fine-tuning our models on the CALLHOME adaptation set (as done in ~\cite{fujita2019end, horiguchi2020end}).
% However, since we did not fine-tune the separation networks on the CALLHOME development set (as done in ~\cite{horiguchi2020end, kinoshita2022tight}), it failed to outperform these methods in terms of DER performance.

\vspace{-0.2cm}

\subsection{CSS Window Analysis}
\label{subsec:res/css_win_eval}

\begin{figure}[ht]
    \begin{subfigure}[b]{0.45\textwidth}
         \includegraphics[width=\textwidth]{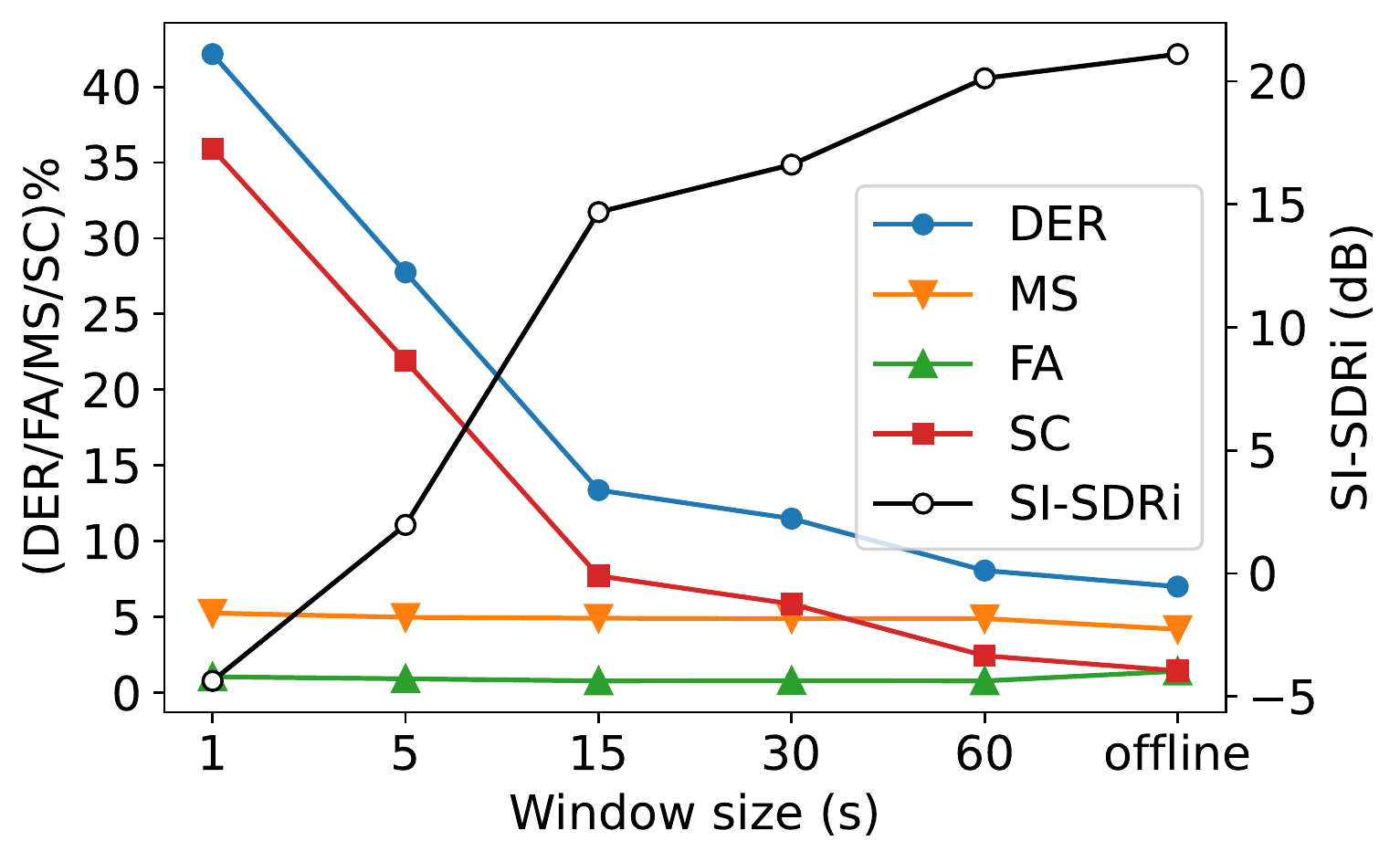}
         \vspace{-0.6cm}
         \caption{Fisher}
         \label{fig:css_analysis/fisher}
     \end{subfigure}
     \vfill
     \begin{subfigure}[b]{0.41\textwidth}
         \includegraphics[width=\textwidth]{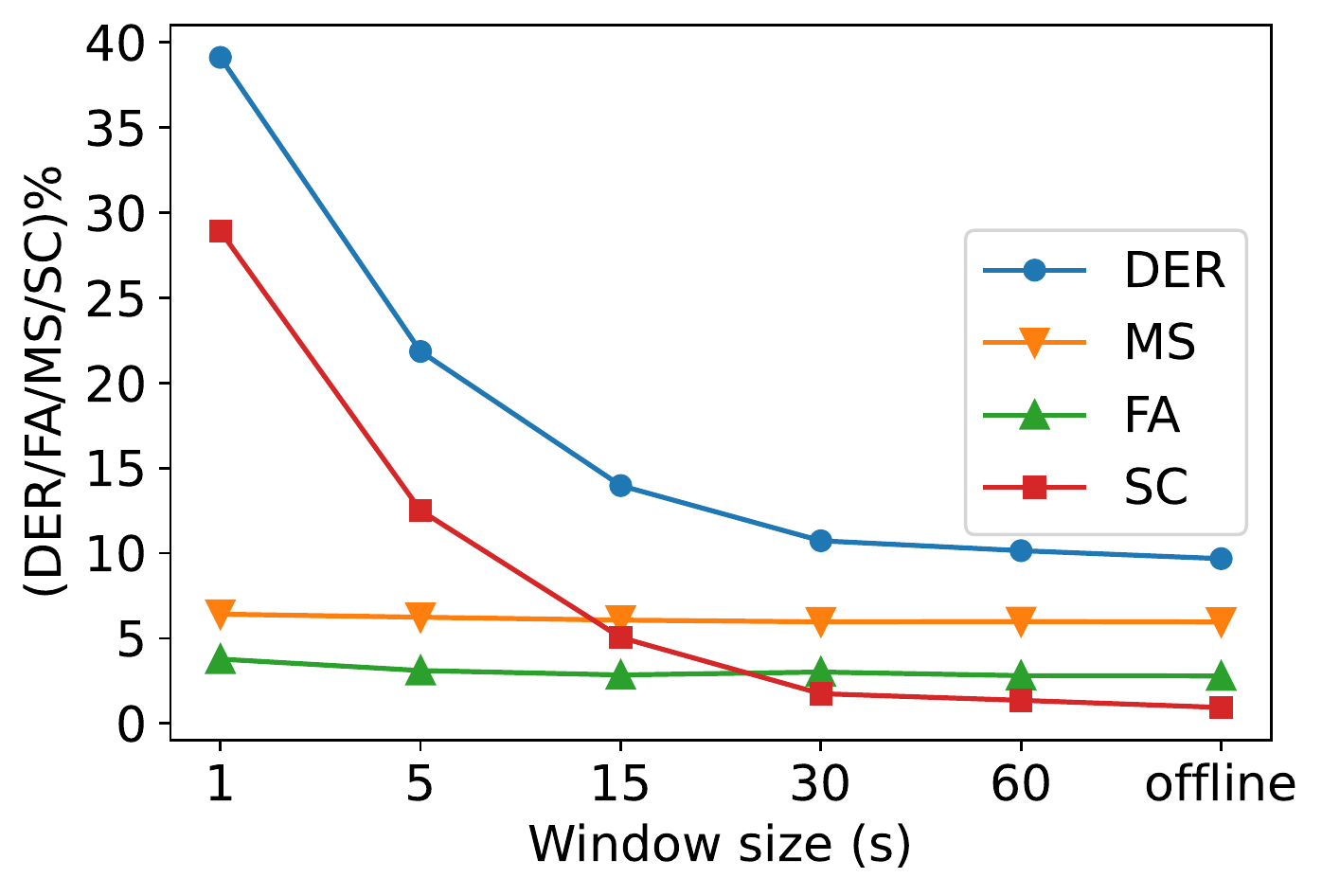}
         \vspace{-0.6cm}
         \caption{CALLHOME}
         \label{fig:css_analysis/callhome}
     \end{subfigure}
     \vspace{-0.3cm}
     \caption{Separation and diarization results on the test sets with different CSS windows. The overlap between windows is set to $50$\%. The results are obtained with the DPRNN+TCN+Leakage removal model.}
     \label{fig:css_analysis}
\end{figure}

Recall from Section \ref{subsec:sys/css} that the CSS framework, besides allowing the processing of arbitrarily long recordings, also allows to use a non-causal separation model in an online fashion with latency reduced to the length of the CSS window. Therefore, it can be regarded as an alternative approach for performing diarization online. 
We use the best SSGD offline model from Table \ref{tab:res/diar_real} (DPRNN+TCN+Leakage removal) to investigate the effect of varying window sizes on SSGD. Evaluation results are reported in Fig.~\ref{fig:css_analysis} for both datasets. As expected, the DER consistently decreased as the window size increased. In particular, the performances were almost on par with the offline models for windows larger than 60 and 30 seconds, respectively, for Fisher and CALLHOME. This suggests a possible parallelization scheme for offline SSGD by applying CSS on minute-long frames simultaneously, resulting in significant inference speed-ups and less memory consumption. The optimal chunk sizes are different for the two datasets because of the difference in their average recording duration (which is 10 minutes and 72 seconds for Fisher and CALLHOME, respectively).

As the window was shortened, missed speech and false alarm error rates remained approximately constant while speaker confusion errors consistently increased, indicating that the main source of error comes from speaker permutation due to wrong channel reordering during the stitching stage of the CSS. For smaller windows, the cross-correlation used for reordering consecutive chunks is less reliable due to the smaller size of the overlapping portion.

The CSS framework is not competitive with the online approach with causal SSep (Sec. \ref{subsec:res/online_eval}) in terms of latency. However, it could be a convenient choice for applications in which better diarization accuracy is more desirable than the low-latency requirement, and memory footprint is an important concern, especially for very long recordings (e.g., $>10$ minutes).

% \begin{figure}
%     \centering
%     \includegraphics[width=.48\textwidth]{offlinedprnn_win.pdf}
%      \caption{Separation and diarization results on the test sets with different CSS windows. The overlap between windows is set to $50$\%. The results are obtained with the DPRNN+TCN+Leakage removal model.}
%      \label{fig:css_analysis}
% \end{figure}

\subsection{Automatic Speech Recognition Evaluation}
\label{subsec:res/wer_eval}

A great advantage of the SSGD framework over other diarization methods is that separated sources together with the segmentation provided by the VAD can be readily fed in input to a back-end ASR system. 
To investigate ASR performance, we feed to a downstream ASR the estimated sources for the DPRNN models with and without leakage removal and using oracle segmentation or not.
We use the pre-trained Kaldi ASPiRE ASR model~\cite{povey2011kaldi}\footnote{\url{https://kaldi-asr.org/models/m1}} and report the performance in terms of word error rate (WER).
We compare the results with the ones obtained with input mixtures and oracle sources, which ideally represent the upper and the lower bound for WER (\%) evaluation.

The results are reported in Table \ref{tab:res/wer_eval}.
We can see that for all SSGD systems the degradation was small compared to using oracle signals. This suggests that the separation is highly effective. 
A large improvement was obtained over the mixture, and we can observe that the main source of performance degradation versus a fully oracle system (oracle VAD + oracle sources) comes from the VAD segmentation. This is consistent with what we observed for diarization in Sections \ref{subsec:res/online_eval} and \ref{subsec:res/offline_eval}.
The leakage removal algorithm slightly degraded the performance, but, on the other hand, in the proposed framework it could be only used for obtaining the segmentation and avoided for ASR (\emph{+ Leakage removal (seg-only)}). In this latter case the performance was slightly increased. 

\begin{table}[t]
\caption{WER evaluation on the Fisher test set. The best online/offline non-oracle results are reported in \textbf{bold}.}
\centering
\adjustbox{max width=\textwidth}{%
\centering
\begin{tabular}{@{}lcccc@{}}
\toprule
\multirow{2}{*}{\textbf{Method}} &
\multirow{2}{*}{\textbf{Online}} &
\multicolumn{2}{c}{\textbf{VAD}} \\
\cmidrule(r{4pt}){3-4}
 &  & \multicolumn{1}{l}{\textbf{TCN}} & \multicolumn{1}{l}{\textbf{Oracle}}\\
\midrule
\textit{Mixture} &  & 38.74  & 30.69 & \\
\textit{Oracle sources} & & 25.44 &  19.50 &  \\
\midrule
DPRNN  & \multicolumn{1}{c}{\cmark} & 26.42 &  20.89 & \\
+ Leakage removal & \multicolumn{1}{c}{\cmark} & 26.94 & 21.03 & \\
+ Leakage removal (seg-only) & \multicolumn{1}{c}{\cmark} & \textbf{26.21} & n.a. & \\
\midrule
DPRNN & \multicolumn{1}{c}{\xmark} & 26.21 &  21.13 & \\
+ Leakage removal & \multicolumn{1}{c}{\xmark} & 26.68 & 21.26 & \\
+ Leakage removal (seg-only) &  \multicolumn{1}{c}{\xmark} & \textbf{26.13} & n.a. & \\
\bottomrule
\end{tabular}
}
\label{tab:res/wer_eval}
\end{table}

% 20.89 dprnn online - oracle
% 30.69 mixture - oralce 
% 19.50 oracle sources - oracle 
% 21.13 dprnn offline - oracle 
%  21.03 dprnn online - leakage rem -oracle
% 21.26 dprnn offline -leakage_rem - oracle 

% 25.44 oracle sources - tcn 
% 26.94 dprnn_online_leakrem - tcn 
% 26.68 dprnn_offline_leakrem - tcn 
% 26.21 dprnn_offline - tcn 
% 26.42 dprnn_online - tcn

\section{Conclusion and Future Work}
\label{sec:conc}

In this paper, we have performed an analysis of SSGD for real-world telephone conversations extending it to online and arbitrarily long diarization scenarios. We have shown that our best online SSGD achieved comparable performances with state-of-the-art methods based on EEND on the CALLHOME dataset with significantly lower latency (for instance, $0.1$ s compared to $10$ s).
Additionally, we have analyzed how the use of CSS with non-causal separation models impacted downstream diarization performance, and have found that DERs were almost on par with the offline case with a sufficiently large CSS window of 60 or 30 seconds for Fisher and CALLHOME datasets, respectively.
Finally, we have shown that SSGD is particularly appealing for multi-talker speaker-attributed ASR, since the estimated sources could be fed directly to an ASR module, leading to significant ASR performance boost. 

These findings open up several research directions.
% We showed that these systems achieve comparable performances with state-of-the-art methods based on clustering or EEND on the Fisher and CALLHOME datasets with significantly lower latencies (for instance, $100$ ms compared to $10$ s). Additionally, for the online scenario, we analyzed the impact of CSS window sizes on downstream diarization performance, and found that we can obtain DERs almost on par with the offline case with a sufficiently large window of 30 or 60 seconds. These findings open up several avenues worth exploring.
First, the gap between the best proposed system and the oracle source evaluation (Table~\ref{tab:res/diar_real}) for Fisher suggested that error mainly came from the VAD module. Future work could investigate joint fine-tuning of separation and VAD to reduce these errors, e.g. on the CALLHOME adaptation set. %Moreover, performance could be improved by training on more data and applying fine-tuning on the CALLHOME adaptation set.
Another direction is to extend the SSGD framework performance to domains other than CTS (e.g., meeting-like and dinner scenarios) where an higher number of speakers could be involved. This could require the development of new techniques since most current source separation methods struggle to track $3$ or more speakers for very long inputs.

%\section{Acknowledgements}
%\label{sec:ack}
\vspace{0.3cm}
\noindent
\textbf{Acknowledgments.} 
This work has been supported by the AGEVOLA project (SIME code 2019.0227), funded by the Fondazione CARITRO.

\clearpage
\bibliographystyle{IEEEbib}

{\footnotesize
\bibliography{refs}}

\end{document}